%% file: main.tex
\documentclass[amsmath,amssymb,twocolumn,prd,floatfix,showpacs, nofootinbib]{revtex4-1}

\usepackage{amsmath} 
\usepackage{graphicx} 
\usepackage{subfigure}
\usepackage{epstopdf} 
\usepackage{color} 
\usepackage{multirow}
\usepackage{setspace} 
\usepackage{overpic} 
\usepackage{amssymb}
\usepackage[normalem]{ulem}
\usepackage{natbib}
\usepackage[bookmarksnumbered, pdfstartview=FitH,colorlinks,urlcolor=blue, citecolor=blue,linkcolor=blue] {hyperref}
\usepackage{lineno}
\usepackage{bm}
\usepackage{rotating}
\usepackage{xcolor}
\usepackage{makecell}
\usepackage{mathtext}
\usepackage{mathrsfs}
\usepackage{graphicx} 
\usepackage{hyperref} 
\usepackage{amsmath} 
\usepackage[utf8]{inputenc}
\usepackage[T1]{fontenc} 
\usepackage{tikz-feynman}  
\usetikzlibrary{arrows.meta}

\let\oldequation\equation
\let\oldendequation\endequation
\renewenvironment{equation}
 {\linenomathNonumbers\oldequation}
 {\oldendequation\endlinenomath}
 
 \newcommand{\BESIIIorcid}[1]{\href{https://orcid.org/#1}{\hspace*{0.1em}\raisebox{-0.45ex}{\includegraphics[width=1em]{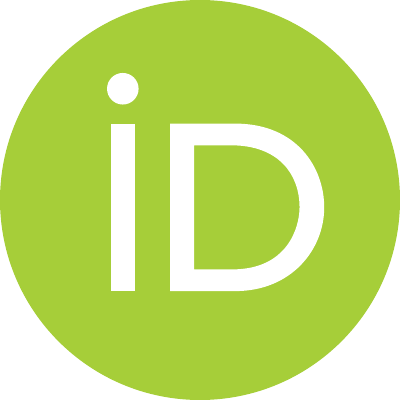}}}}

 \hypersetup{colorlinks = true,
            linkcolor = blue,
            urlcolor = blue,
            citecolor = blue,
            breaklinks=true,
            pdfstartview=Fit}

\begin{document}
\title{\boldmath Measurement of the branching fraction of  $\eta \rightarrow \mu^+ \mu^-$ and  search for $\eta \rightarrow e^+ e^-$}

\author{
\begin{center}
\input{authorlist}
\end{center}
\vspace{0.4cm}
}


\begin{abstract}
We analyze the decay of $\eta\rightarrow \ell^+\ell^-(\ell=e, \mu)$ via $J/\psi\rightarrow\gamma\eta'$ and $\eta'\rightarrow\pi^+\pi^-\eta$, based on  (10087 $\pm$ 44) $\times$ 10$^{6}$ $J/\psi$ events collected with the BESIII detector at the BEPCII storage rings. The branching fraction of $\eta\rightarrow\mu^+ \mu^-$ is measured to be $(5.8 \pm 1.0_{\rm stat} \pm 0.2_{\rm syst}) \times 10^{-6}$, which is consistent with the previous measurements and theoretical expectations. In addition, no significant $\eta\to e^+e^-$ signal is observed in the $e^+ e^-$ invariant mass spectrum, and an improved upper limit of ${\cal B}(\eta \to e^+ e^-) < 2.2 \times 10^{-7}$ is set at 90\% confidence level.
\end{abstract}

\maketitle

\section{Introduction}
\begin{figure}[htbp]
\centering
\subfigure{\includegraphics[scale=0.55]{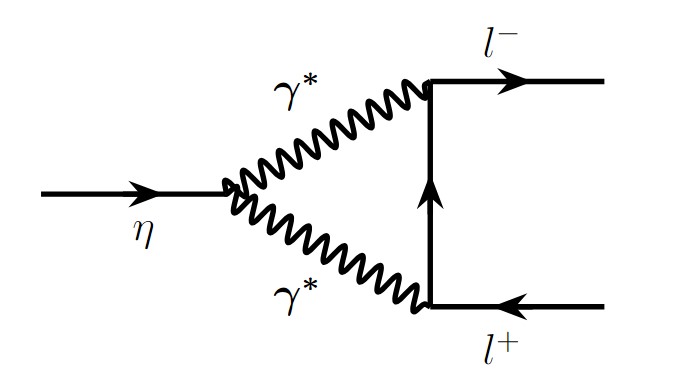}}
\caption{The Feynman diagram for $\eta \rightarrow \ell^+ \ell^-$.}
\label{Feiman ll}
\end{figure}

Within the Standard Model (SM), the decay of $\eta$ into lepton pairs ($\eta\to \ell^+\ell^-$ with $\ell= e$ or $\mu$) proceeds via a fourth-order electromagnetic transition. The process is dominated by a two-photon intermediate state, as depicted in Fig.~\ref{Feiman ll}, resulting in significant suppression of the decay. Especially for $\eta\rightarrow e^+e^-$, the decay probability is expected to be very low, since the helicity factor is proportional to the square of the electron mass. Hence, a measurement of this decay probability higher than the SM model prediction could be indicative of hypothetical interactions arising from physics beyond the SM. This has stimulated considerable theoretical effort~\cite{Young:1967zzb,BABU1982449,AMETLLER1983301,PhysRevD.29.911, PhysRevD.42.1509, gengLeptonicCPViolation1990, Margolis:1992gf,SAVAGE1992481,PhysRevLett.80.4633,Ametller:1993we,Knecht:1999gb,Silagadze:2006rt,Dorokhov:2009xs,Petri:2010ea}  devoted to calculating the decay rate of $\eta\to \ell^+\ell^-$. The predicted branching fractions for $\eta\to\mu^+\mu^-$ and $\eta\to e^+ e^-$, based on the Canterbury approximants method~\cite{Masjuan_2016}, are $(4.72^{+0.05}_{-0.21})\times10^{-6}$ and $(5.31^{+0.14}_{-0.04})\times10^{-9}$,
respectively. 

Experimentally, a stringent upper limit of $\mathcal{B}(\eta\rightarrow e^+ e^-) < 7\times 10^{-7}$, at 90\% confidence level (CL), was obtained from the inverse process $e^+e^-\to\eta$~\cite{SND:2018egu}. The world average of $\mathcal{B}(\eta\rightarrow\mu^{+}\mu^{-})=(5.7\pm0.8 )\times 10^{-6}$~\cite{PhysRevD.110.030001} is a combination of measurements performed by \mbox{SATURNE} II ~\cite{PhysRevD.50.92} and  Lepton-G~\cite{DZHELYADIN1980471} more than thirty years ago.
In recent years, taking advantage of the abundant production of $\eta, \eta^\prime$ mesons in $J/\psi$ radiative decays and the high statistics $J/\psi$ events, the BESIII experiment has made significant improvements in the study of their decays. However, the investigation of $\eta\to\ell^+\ell^-$ decays via the radiative decay $J/\psi\to\gamma\eta$ suffers from large backgrounds from $J/\psi\to\gamma\ell^+\ell^-$decays. 

More recently, a novel approach~\cite{Kang:2023yye} was proposed to  investigate  $\eta$ decays, particularly for their rare or forbidden decays, with  $\eta^\prime\rightarrow\pi\pi\eta$ produced in $J/\psi$ radiative decays.   
Therefore, using the sample of  $(10087 \pm 44) \times$ 10$^{6}$~\cite{BESIII:2021cxx} $J/\psi$ events collected with the BESIII detector, we perform a study of $\eta\rightarrow \ell^+\ell^-$ with $J/\psi\rightarrow\gamma\eta'$ and the subsequent decay of $\eta'\rightarrow\pi^+\pi^-\eta$.

\section{BESIII DETECTOR and MONTE CARLO SIMULATION}
\label{sec:BES}
The BESIII detector~\cite{ABLIKIM2010345} is a magnetic spectrometer  located at the Beijing Electron Positron Collider (BEPCII)~\cite{ABLIKIM2010345}. The cylindrical core of the BESIII detector covers 93\% of the full solid angle and consists of a helium-based multilayer drift chamber (MDC), a plastic scintillator time-of-flight system (TOF), and a CsI(Tl) electromagnetic calorimeter (EMC), which are all enclosed in a superconducting solenoidal magnet providing a 1.0~T magnetic field. The  magnetic field was 0.9~T in 2012, which affects 11\% of the total $J/\psi$ data. The  solenoid is supported by an octagonal flux-return yoke with resistive plate counter muon identifier modules interleaved with steel. The charged-particle momentum resolution at 1~GeV/$c$ is 0.5\%, and the  specific ionization energy loss $\text{d}E/\text{d}x$ resolution is 6\% for the electrons from Bhabha scattering. The EMC measures photon energies with a resolution of 2.5\% (5\%) at 1~GeV in the barrel (end-cap) region. The time resolution of the 
TOF barrel part is 68~ps, while that in the end-cap part is 110~ps. The end-cap TOF system was upgraded in 2015 using multi-gap resistive plate chamber technology, providing a time resolution of 60~ps~\cite{Li:2017jpg,Guo:2017sjt,CAO2020163053}, which benefits 87\% of the data used in this paper.

Monte Carlo (MC) simulated data samples are produced with a {\sc geant4}-based~\cite{AGOSTINELLI2003250} software package, which includes the geometric description of the BESIII detector and the detector response. These samples are used to determine detection efficiencies and  estimate the backgrounds.  To thoroughly investigate potential backgrounds, we utilize an inclusive MC sample comprising 10 billion $J/\psi$ events.  The inclusive MC sample includes both the production of the $J/\psi$ resonance and the continuum processes incorporated in  {\sc kkmc}~\cite{JADACH2000260,PhysRevD.63.113009}. All particle decays are modeled with {\sc evtgen}~\cite{LANGE2001152} using the branching fractions either taken from the Particle Data Group (PDG)~\cite{PhysRevD.110.030001}, when available, or otherwise estimated with {\sc lundcharm}~\cite{PhysRevD.62.034003}. Final state radiation (FSR) from charged final state particles is incorporated using the {\sc photos} package~\cite{BARBERIO1991115}. Specific generators using the appropriate theoretical amplitudes for $\eta'\rightarrow\pi^+\pi^- \ell^+ \ell^-$~\cite{Zhang:2012gt}, $\eta'\rightarrow\pi^+\pi^- \pi^+ \pi^-$~\cite{BESIII:2014bgm}, and $\eta\rightarrow\gamma e^+ e^-$~\cite{Qin:2017vkw} decays were developed for this analysis.
 
\section{EVENT SELECTION}
\label{sec:selection}
The final state of interest is studied through the decay chain $J/\psi\rightarrow\gamma\eta', \eta'\rightarrow\pi^+\pi^-\eta$, and  $\eta\rightarrow \ell^+ \ell^-$. Each event is required to contain at least one good photon candidate and four charged track candidates with a total charge of zero. Each charged track reconstructed from MDC hits is required to have a polar angle in the range $|\cos\theta|\leq0.93$ and to pass the interaction point within $\pm10$ cm along the beam direction and within $\pm1$ cm in the plane perpendicular to the beam.

Photon candidates are reconstructed using isolated clusters of energy deposited in the EMC and must have a deposited energy larger than 25 MeV in the barrel region
($|\cos\theta|\leq0.80$) and 50 MeV in the end-cap region ($0.86<|\cos\theta|<0.92$). The energy deposited in nearby TOF counters is included to improve the reconstruction
efficiency and energy resolution. To eliminate clusters associated with charged tracks, the angle between the photon candidate and the extrapolation of any charged track to the EMC must be larger than 10 degrees. A requirement on the EMC cluster timing with respect to the event start time
$(0 \leq T \leq 700 \, \text{ns})$
is used to suppress electronic noise and
energy deposits unrelated to the event.



For each candidate, particle identification (PID) is performed using the TOF and $\text{d}E/\text{d}x$ information, and four-constraint (4C) kinematic fits are performed imposing energy and momentum conservation under the hypotheses of $J/\psi\to\gamma\pi^+\pi^- \ell^+ \ell^-$. For each event, the combination with the smallest $\chi_{\text{sum}}^2 (\gamma\pi^+\pi^-\ell^+\ell^-)$ is kept for further analysis, where $\chi_{\text{sum}}^2(\gamma\pi^+\pi^-\ell^+ \ell^-) = \chi_{\text{4C}}^2 + \sum_{i=1}^4 \chi_{\text{PID}(i)}^2$ is the sum of the 4C kinematic fit $\chi^2$ and PID $\chi^2$ of the four charged tracks.


\section {Analysis of $\eta\rightarrow\mu^+\mu^-$}
\label{Sec:BR_determined}

 




For the $\eta\rightarrow\mu^+\mu^-$ decay, the events with  $\chi^{2}_{\rm sum}(\gamma\pi^+\pi^- \mu^+ \mu^-)<40$ are retained. Here the requirement is  optimized by  the figure of merit
\cite{Punzi:2003bu} $S/\sqrt{S+B}$, where $S$ is the number of signal events from MC simulation and $B$ is the number of background events in data. To reject possible background events from  $J/\psi\rightarrow\gamma 2(\pi^+\pi^-)$, $\chi^2_{\rm sum} (\gamma \pi^+ \pi^- \pi ^+ \pi ^-)> \chi^2_{\rm sum} (\gamma\pi^+ \pi^- \mu^+ \mu^-)$ is required. The muon counter information is not further used due to the low momentum of the muons of interest.

\begin{figure}[htbp]
    \centering    \includegraphics[width=0.9\columnwidth]{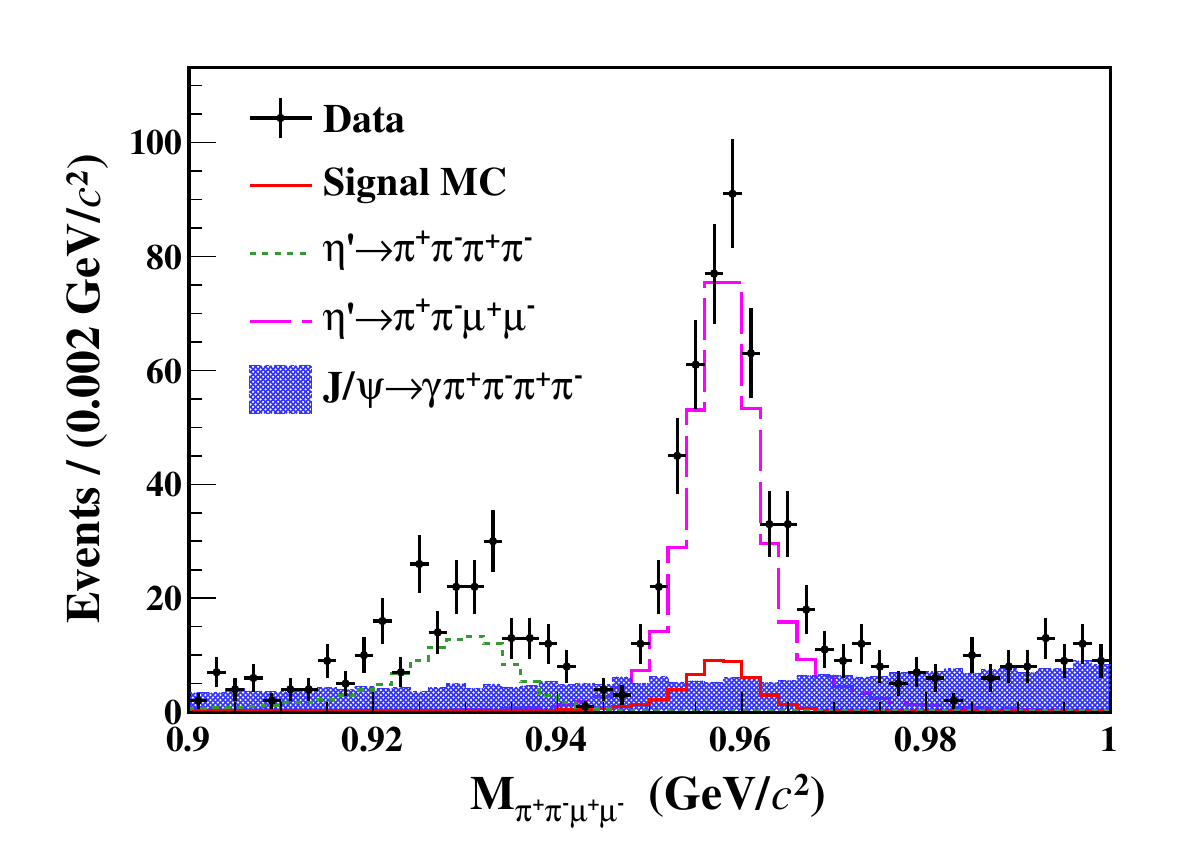}
    \caption{ The distribution of $M_{\pi^+\pi^-\mu^+\mu^-}$.  The dots with error bars represent the data, and the red histogram is the signal MC  sample. The dashed green line, dashed pink line, and shaded blue histogram represent the backgrounds from $\eta\rightarrow\pi^+\pi^-\pi^+\pi^-$, $\eta'\rightarrow\pi^+\pi^-\mu^+\mu^- $ and $J/\psi\rightarrow\gamma\pi^+\pi^-\pi^+\pi^-$, respectively.}
    \label{Mpipimumu}
\end{figure}
\begin{figure}[htbp]
    \centering    \includegraphics[width=0.9\columnwidth]{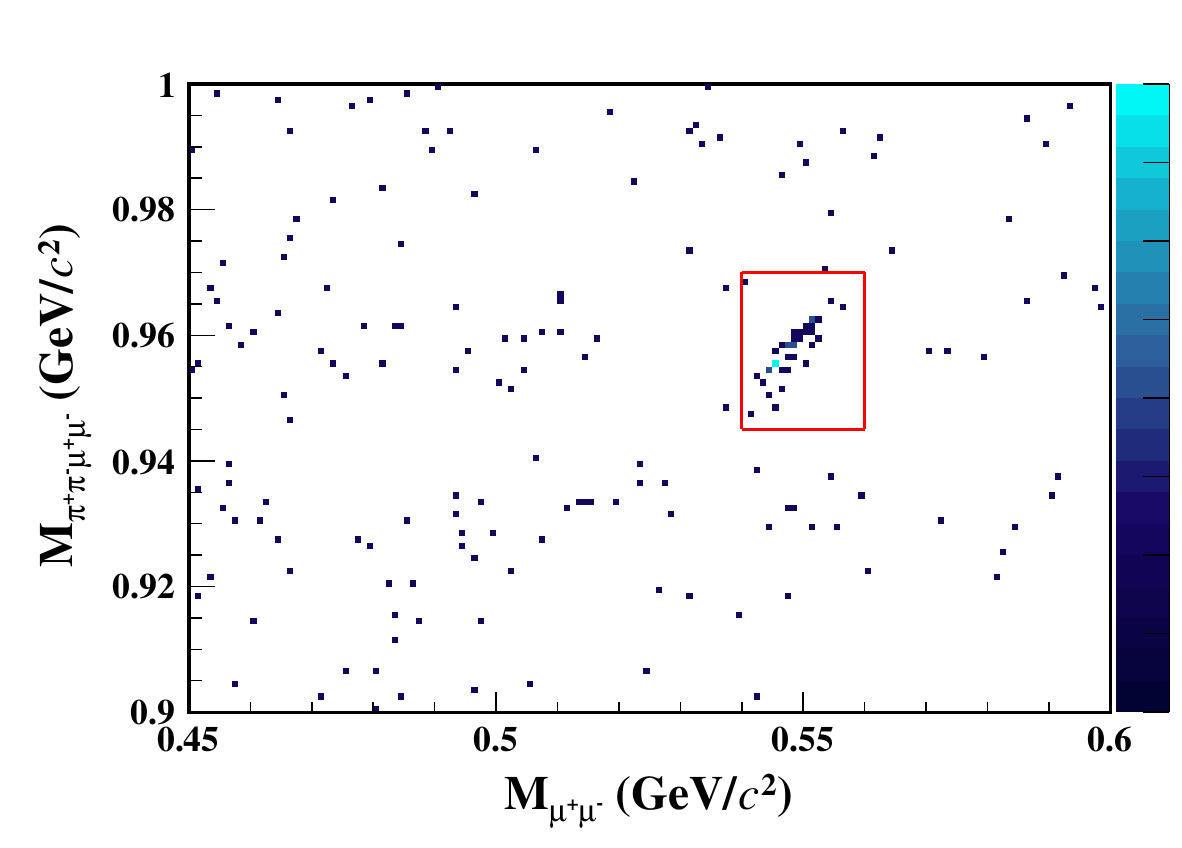}
    \caption{ The distribution of $M_{\pi^+\pi^-\mu^+\mu^-}$  versus $M_{\mu^+\mu^-}$ for data.}
    \label{Plot_Mmm_ppmm}
\end{figure}
\begin{figure}[htbp]
    \center
 \includegraphics[width=0.9\columnwidth]{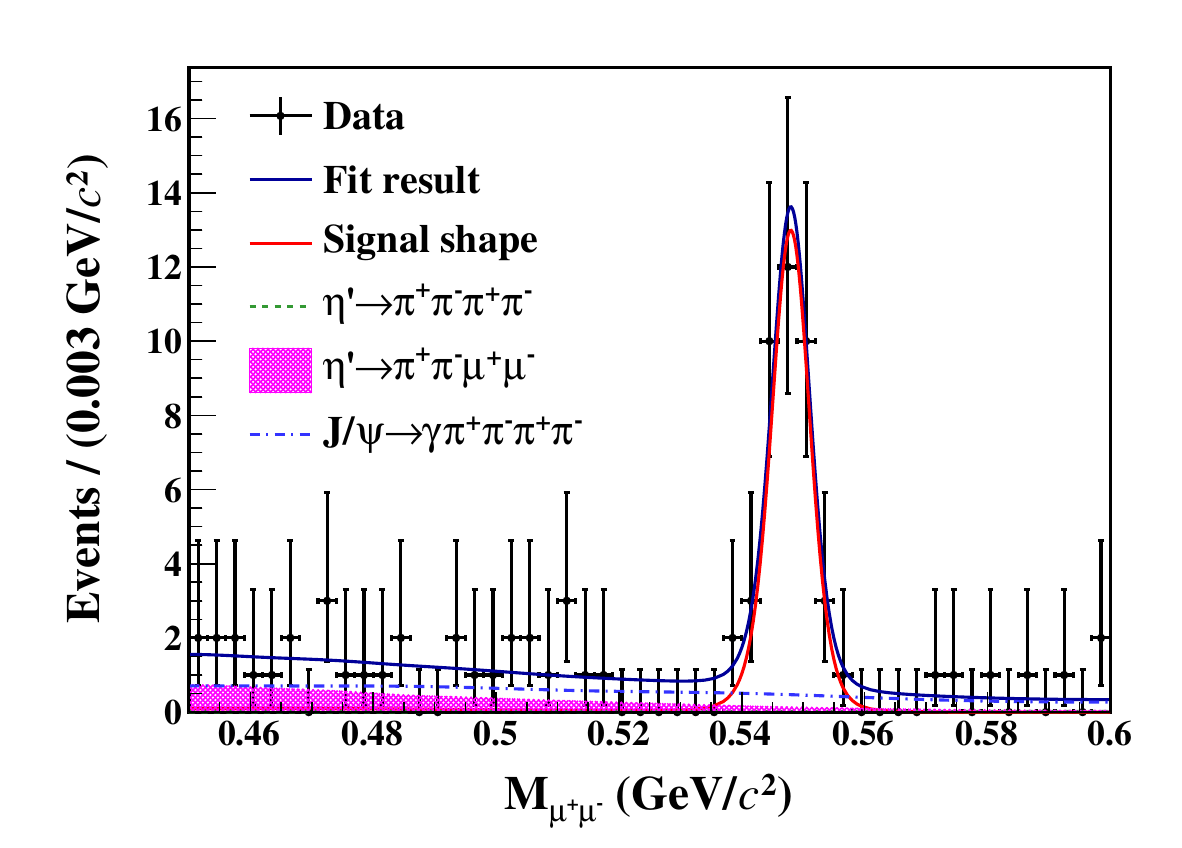}
    \caption{The fit to the $M_{\mu^+\mu^-}$ distribution. The dots with error bars represent the data. The blue line represents the total fit result, and the red line represents the signal component. The dashed green line, shaded pink shape, and dot-dash blue line are the contributions from the decays $\eta'\rightarrow\pi^+\pi^-\pi^+\pi^-$,    $\eta'\rightarrow\pi^+\pi^-\mu^+\mu^-$ and $J/\psi\rightarrow\gamma\pi^+\pi^-\pi^+\pi^-$, respectively.}

    \label{fit_result}
\end{figure}

 The resulting $\pi^+\pi^-\mu^+\mu^-$  invariant mass distribution ($M_{\pi^+\pi^-\mu^+\mu^-}$) is shown in Fig.~\ref{Mpipimumu}. A clear $\eta^\prime$ peak is observed, while the small peak around 0.93 $\text{GeV}/c^2$ comes from the background $\eta'\rightarrow\pi^+\pi^-\pi^+\pi^-$ decay. The distribution of  $M_{\pi^+\pi^-\mu^+\mu^-}$ versus the $\mu^+\mu^-$  invariant mass ($M_{\mu^+\mu^-}$) is displayed in Fig.~\ref{Plot_Mmm_ppmm}.  The cluster within the marked region corresponds to the $\eta^\prime\to\pi^+\pi^-\eta$ decay.  Requiring $M_{\pi^+\pi^-\mu^+\mu^-} \in (0.945, 0.970)~\text{GeV}/c^2$, an evident $\eta$ peak is seen in the $M_{\mu^+\mu^-}$ distribution, as shown in Fig.~\ref{fit_result}. Although the $\eta'$ peak in Fig.~\ref{Mpipimumu} is dominated by background events from $\eta'\rightarrow\pi^+\pi^-\mu^+\mu^-$, these events are  primarily concentrated in the low-energy region of $M_{\mu^+\mu^-}$ and contribute very little in the mass range of interest, as shown in Fig.~\ref{fit_result}.

To investigate background events in  the $M_{\mu^+\mu^-}$ distribution,  an inclusive MC sample of 10 billion $J/\psi$ events is analyzed with the same procedure. The background events are mainly from the decays $J/\psi\rightarrow\gamma\eta',\eta'\rightarrow\pi^{+}\pi^{-}\pi^{+}\pi^{-}$, $J/\psi\rightarrow\gamma\eta',\eta'\rightarrow\pi^{+}\pi^{-}\mu^{+}\mu^{-}$  and $J/\psi\rightarrow\gamma\pi^{+}\pi^{-}\pi^{+}\pi^{-}$. Dedicated MC samples for these background channels are generated to estimate their contributions. We find that none of these possible background channels contributes to the $\eta$ peak, as seen in Fig.~\ref{fit_result}.

To obtain the signal yield of $\eta\rightarrow\mu^+\mu^-$, an extended unbinned maximum likelihood fit is performed to the  $M_{\mu^+\mu^-}$ distribution.  In the fit, the total probability density function
consists of a signal component and various background contributions. The signal component is modeled with the simulated shape. The background components considered are subdivided into two classes. For the first, the shapes of the background events from $\eta^\prime$ decays $J/\psi\rightarrow\gamma\eta',\eta'\rightarrow\pi^+\pi^-\mu^+\mu^-$ and $J/\psi\rightarrow\gamma\eta',\eta'\rightarrow\pi^+\pi^-\pi^+\pi^-$ are taken from  dedicated MC simulations. The magnitudes of these backgrounds are fixed at $15.0\pm3.0$ and $0.8\pm0.1$, respectively, normalized using the branching fractions of $J/\psi\rightarrow\gamma\eta'$ and their subsequent decays.  The second class is the $J/\psi\rightarrow \gamma \pi^+\pi^-\pi^+\pi^-$ component, also described with the MC-simulated shape, with a magnitude left as a free parameter in the fit.

The fit, shown in  Fig.~\ref{fit_result}, gives  37.9 $\pm$ 6.7  
 $\eta\rightarrow \mu^+\mu^-$ events with a statistical significance of 9.8$\sigma$. This significance is calculated from the
 change in the negative log likelihood function $-\text{ln}\mathcal{L}$ with and without the signal component, accounting for the change in degrees of freedom between the two fit configurations. Taking into account the detection efficiency of (29.19 $\pm$ 0.07)\% obtained from the MC simulation, the branching fraction of $\eta\rightarrow\mu^+\mu^-$is measured to be
\begin{equation}
\begin{aligned}
\mathcal{B}(\eta\rightarrow\mu^+\mu^-) & \,=\, 
 \frac{N_{\rm obs}}
      {N_{J/\psi}\cdot  {\mathcal B}_{\rm int} \cdot \varepsilon_{\eta\rightarrow\mu^+\mu^-}} \\
\\
    & \,=\, (5.8 \pm 1.0)  \times 10^{-6}.
\end{aligned}
\end{equation}
where  $N_{\rm obs}$ is the signal yield, $N_{J/\psi}$ is the total number of $J/\psi$ events, $\mathcal{B}_{\rm int}\equiv \mathcal{B}(J/\psi\rightarrow\gamma\eta') \, \mathcal{B}(\eta'\rightarrow\pi^+\pi^-\eta)$ is taken from the PDG~\cite{PhysRevD.110.030001},  $\varepsilon_{\eta\rightarrow\mu^+\mu^-}$ is the detection efficiency, and the uncertainty is statistical only. 

\section{Analysis of $\eta\rightarrow e^+ e^-$}
\begin{figure}[htbp]
    \centering
\includegraphics[width=0.9\columnwidth]{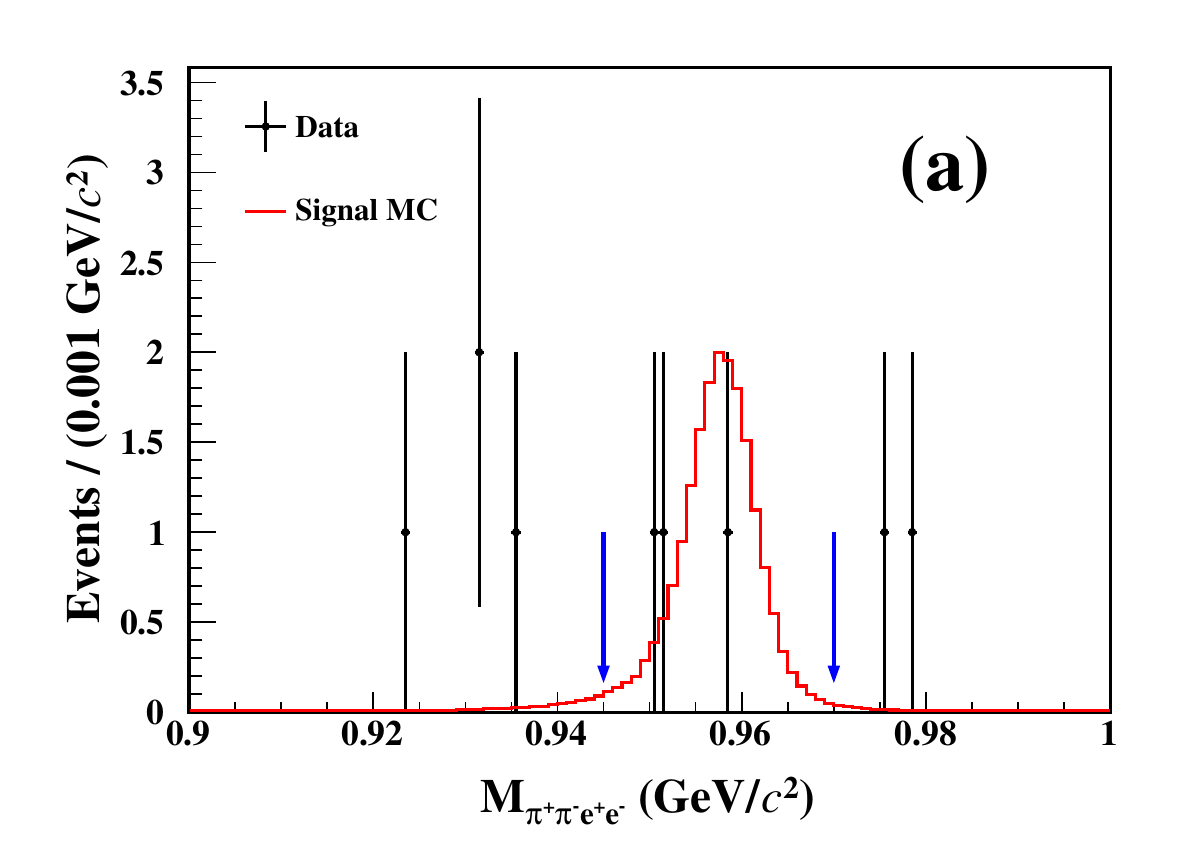}     \includegraphics[width=0.9\columnwidth]{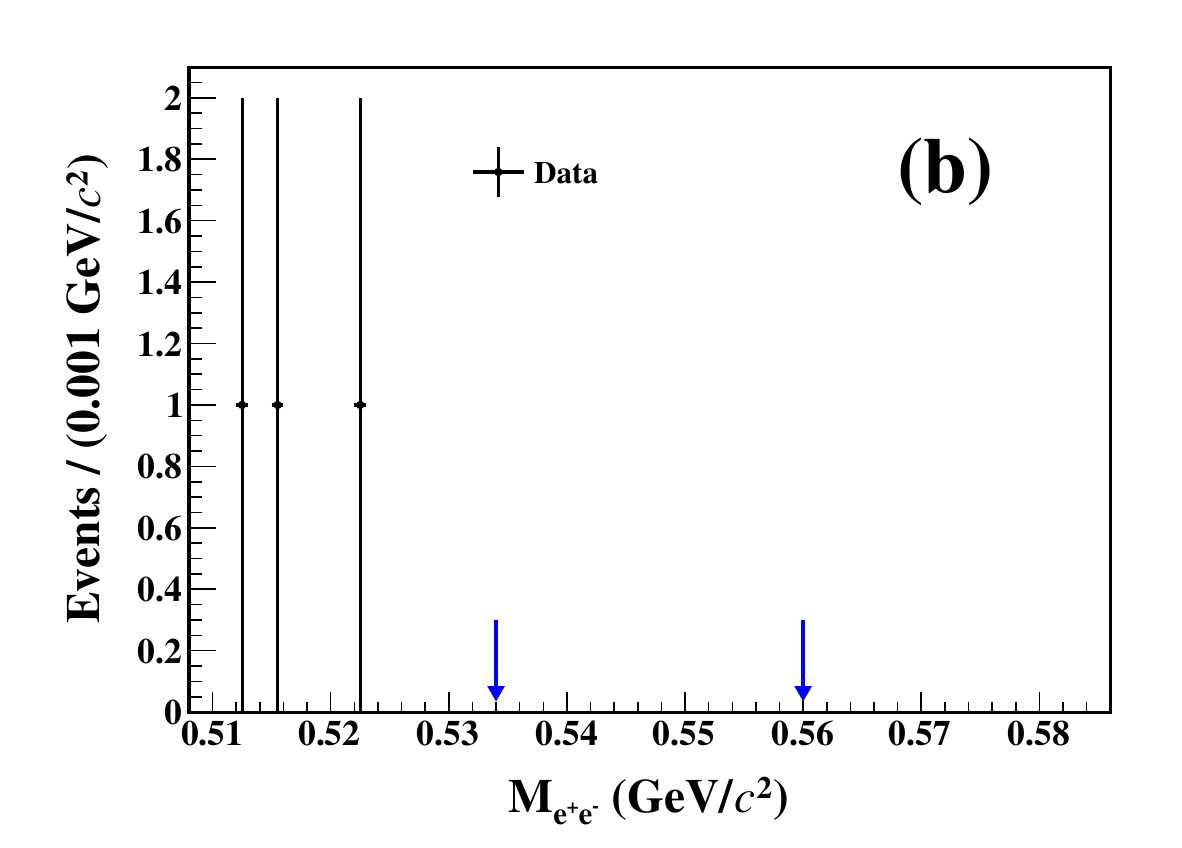} 
    \caption{ The distributions of (a) $M_{\pi^+\pi^- e^+ e^-}$ and (b) $M_{e^+ e^-}$. Dots with error bars represent the data, the red  histogram is the signal MC shape. Two blue arrows represent the signal region.}
    \label{Mee}
\end{figure}

To select $\eta\to e^+ e^-$ events, the  $\chi^{2}_{\rm sum}(\gamma\pi^+\pi^-e^+e^-)$ is required to be less than 40.  The $\pi^+\pi^- e^+ e^-$ invariant mass ($M_{\pi^+\pi^- e^+ e^-}$) is shown in  Fig.~\ref{Mee}(a). No evident $\eta^\prime$ signal is observed. We then select the $\eta^\prime$ mass window to be within $(0.945, 0.97) \,\text{GeV}/c^2$.  Fig.~\ref{Mee}(b) shows the $e^+e^-$ invariant mass ($M_{e^+e^-}$) distribution for candidates passing all selection criteria. No events are observed in the $\eta$ mass region.

A background study based on the inclusive MC sample indicates that the dominant background originates from $J/\psi\rightarrow\gamma\eta'$ with the subsequent decays $\eta'\rightarrow\pi^+\pi^-\eta$, $\eta\rightarrow\gamma e^+e^-$ and $\eta'\rightarrow\pi^+\pi^- e^+e^-$. Dedicated MC samples are generated to evaluate contributions from these channels. The background yields are  $0.5\pm0.3$ and $1.0\pm0.2$ events, respectively.

The signal region is defined as the interval within 3.5 standard deviations. The standard deviation represents the mass resolution obtained from signal MC events, corresponding to a mass interval of $(0.534, 0.560)\, \text{GeV}/c^2$.  
No events are observed in this interval, $N_{\rm sig}=0$. The background contribution, $N_{\rm bkg}$, in the signal region is estimated using events in sideband regions. 
Symmetric sideband regions, each with the same width as the signal region, are defined as $(0.508, 0.534) \, \text{GeV}/c^2$ and $ (0.560, 0.586) \, \text{GeV}/c^2$, giving $N_{\rm bkg} = 1.5$. With the dedicated MC sample, the detection efficiency is determined to be  $(27.48\pm0.04)\%$, where the uncertainty is statistical only.

The upper limit on the signal yield, \( N_{\rm UL} \), is estimated  at a CL of 90\% using the  method from Ref.~\cite{Rolke:2000ij}, which is suitable for small signals in the presence of background.  Given \( N_{\rm sig} \) and \( N_{\rm bkg} \), \( N_{\rm UL} \) of $\eta\to e^+e^-$ is  estimated to be 1.3. Accounting for efficiency and systematic uncertainty $ \delta_{\rm sys}$ (see Section~\ref{system} below for details), the upper limit on the branching fraction of $\eta\to e^+e^-$ is determined as
\begin{equation}
\begin{aligned}
    \mathcal{B}(\eta\rightarrow e^+ e^-) & \,<\, 
  \frac{N_{\rm UL}}
       {N_{J/\psi} \cdot \mathcal{B}_{\rm int} \cdot \varepsilon_{\eta \to e^+ e^-}\cdot(1 - \delta_{\rm sys})}\\
    \\
&= 2.2 \times 10^{-7}.
\end{aligned}    
\end{equation}

\section{Systematic Uncertainties}
\label{system}
The sources of systematic uncertainties include the efficiency differences between the data the and MC simulation for the MDC tracking, the PID, the photon detection, and the
kinematic fit,  as well as the fitting procedure and the total number of $J/\psi$ events. Their effects
on the branching fractions are discussed below.

\begin{itemize}
    \item  Total number of $J/\psi$ events:
    
The total number of $J/\psi$ events, (10087 $\pm$ 44) $\times$ 10$^{6}$~\cite{BESIII:2021cxx}, is determined from an analysis
of inclusive $J/\psi$ hadronic events, and the uncertainty, 0.4\%, is taken as the systematic uncertainty.

    \item Uncertainties of $\mathcal{B}(J/\psi\rightarrow\gamma\eta')$ and $\mathcal{B}(\eta'\rightarrow\pi^+ \pi^- \eta)$:
   
    The branching fractions of $J/\psi\rightarrow\gamma\eta'$ and $\eta'\rightarrow\pi^+ \pi^- \eta$ are
    taken from the PDG~\cite{PhysRevD.110.030001}. 
    Their uncertainties are 1.3\% and 1.2\%, respectively.

  \item MDC tracking:
   
    The single charged pion track efficiency has been studied with respect to the transverse momentum and polar angle $\theta$ in the MDC  by means of a high-purity sample of  $J/\psi\rightarrow \pi^+\pi^-\pi^0$~\cite{liuStudyTrackingEfficiency2025} obtained by using one pion to tag the presence of the other.
   Two-dimensional tracking efficiencies as a function of the transverse momentum and $\cos\theta$ are evaluated for both the data and MC simulation. To correct the MC efficiencies to those of the data, the corresponding two-dimensional correction factors and their uncertainties are derived.  To estimate its impact on the systematic uncertainty, the MC simulation events are then weighted to determine the detection efficiency. Since the mass and the behavior of low-momentum muons are similar to those of pions, we weight the muon tracks by assuming them to be pions.
    
In the case of electrons, the clean samples of $e^+ e^-\rightarrow e^+ e^-\gamma$ at the $J/\psi$ meson mass and $J/\psi\rightarrow e^+ e^-\gamma_{\rm FSR}$ are used to evaluate the MC-data tracking difference as a function of momentum and polar angle, which are then used to weight the MC simulated events.  The changes in the detection efficiencies with and without correction, 0.8\% for  $\eta\rightarrow\mu^+\mu^-$ and 2.1\% for $\eta\rightarrow e^+ e^-$,  are considered as the systematic uncertainties.

   \item PID and 4C kinematic fit:
   
   In case of $\eta\rightarrow\mu^+\mu^-$, the uncertainties from PID and the kinematic fit are evaluated separately.  To estimate the pion PID efficiency uncertainty, we have studied the pion PID efficiency using the control samples of $J/\psi\rightarrow\pi^+\pi^-\pi^0$. The results show that the pion PID efficiency for data agrees well with that of the MC simulation. The minor MC-data discrepancy as a function of the momentum and the polar angle is used to estimate the systematic uncertainty with the same approach as for the tracking efficiency.  Due to a lack of a specific control sample for low-momentum muons, the PID efficiency correction is performed with the correction factor obtained from pions. The uncertainty associated with the 4C kinematic fit is studied by correcting the track helix parameters~\cite{PhysRevD.87.012002} to reduce the difference between data and MC simulation. The differences between the efficiencies with and without corrections, summarized in Table~\ref{eemumusys}, are taken as the systematic uncertainties from PID and 4C kinematic fit.
      
For $\eta\rightarrow e^+ e^-$, we use a clean control sample of $J/\psi\rightarrow \pi^+\pi^-\pi^0,\pi^0\rightarrow\gamma e^+ e^-$ to study the uncertainty. This control sample has the same topology as the decay of interest. Using the same charged tracks and photon selection criteria, we first select the candidate events with a very loose requirement on $\chi^{2}_{ \rm sum}$. Then, we further tighten the selection by requiring  $\chi^{2}_{\rm sum}<40$. The efficiency difference between data and MC simulation, 4.6\%, is taken as the systematic uncertainty. 


Although the systematic uncertainty from PID and the 4C kinematic 
fit is larger than other individual contributions, its combination 
with the tracking efficiency correction yields a small net effect 
on the detection efficiency. We therefore use the result without 
the tracking efficiency correction as the nominal value and take 
the difference between the results with and without the correction 
as the systematic uncertainty.



\begin{table}[htbp]
    \centering
    \caption{Relative systematic uncertainties (in \%) in the branching fraction measurements.}
    \begin{tabular}{lcc}
        \hline\hline
        Source & $\eta\rightarrow\mu^+\mu^-$ &  $\eta\rightarrow e^+ e^-$\\
        \hline
        Number of $J/\psi$ events & $0.4$ & $0.4$  \\
        $\mathcal{B}(J/\psi\rightarrow\gamma\eta')$ & $1.3$ &  $1.3$ \\
        $\mathcal{B}(\eta'\rightarrow\pi^+\pi^-\eta)$ & $1.2$ & $1.2$\\
        MDC tracking           & $0.8$  & $2.1$  \\
        PID and kinematic fit & 2.7  & $4.6$ \\
        Photon detection       & $0.5$  & $0.5$   \\
        Signal shape           & $0.1$ & -       \\
        Background shape       & $0.2$  & -       \\
        \hline
        Total & 3.4 & 5.6 \\
        \hline\hline
    \end{tabular}
    \label{eemumusys}
\end{table}

    \item Photon detection efficiency:
    
    The systematic uncertainty from the reconstruction of photons has been  extensively  studied in  the processes $e^+e^-\rightarrow\gamma\mu^+\mu^-$. The relative difference in efficiency between data and MC simulation, up to 0.5\%, is assigned as the systematic uncertainty. 

    \item Signal shape:
    
    The uncertainties due to the signal MC shape is estimated by performing an alternative fit  in which 
    the signal is modeled by the MC shape convolved with a single Gaussian function. It is found that the difference of the determined number of signal events, 0.1\%, which is taken as the systematic uncertainty. 
       
    \item Background shape:
    
    In the fit to $M_{\mu^+\mu^-}$, the numbers  of background events from   $ \eta'\rightarrow\pi^+\pi^-\pi^+\pi^-$ and $ \eta'\rightarrow\pi^+\pi^-\mu^+\mu^-$ are fixed at the normalized values according to the branching fractions from the PDG~\cite{PhysRevD.110.030001}. To estimate the effect of the uncertainty of the used branching fractions, the numbers of events for these two backgrounds are varied randomly within the uncertainties of the branching fractions.  Using these randomly scaled parameters, we perform a series of fits to the $M_{\mu+\mu^-}$ distribution. The difference of the determined number of signal events, 0.2\%, is taken as the systematic uncertainty.
    
    \item Fit range:

    The systematic uncertainty due to the $\eta$ fit range is evaluated by varying the nominal range by $\pm 0.01$ or $\pm 0.02$ GeV/$c^2$ in six variations.  For each variation, the deviation between the alternative and nominal one is defined as
 $\zeta = \frac{|{\mathcal B}_{\rm nominal} - {\mathcal B}_{\rm test}|}{\sqrt{|\sigma_{{\mathcal B},{\rm nominal}}^2 - \sigma_{{\mathcal B},{\rm test}}^2|}}$, where ${\mathcal B}$ denotes the branching fraction of $\eta\to\mu^{+}\mu^{-}$ and $\sigma$ denotes the statistical uncertainty. According to the method discussed in Ref.~\cite{barlow2002systematicerrorsfactsfictions}, this uncertainty source is negligible because all $\zeta$ values are below 1$\sigma$. 
    
\end{itemize}

Table~\ref{eemumusys} summarizes all contributions to the
systematic uncertainties on the branching fraction measurements. In each case, the total systematic uncertainty is obtained by adding the individual contributions in quadrature.

\section{Summary}

With a sample of (10087 $\pm$ 44) $\times$ 10$^6$ ~\cite{BESIII:2021cxx} $J/\psi$ events collected with the BESIII detector,  we perform a study of $\eta\to \ell^+\ell^-$ tagging with $\eta^\prime\to\pi^+\pi^-\eta$ in $J/\psi$ radiative decays.
The decay $\eta\rightarrow\mu^+\mu^-$ is observed  with a significance of $9.8\sigma$, and its  branching fraction is measured to be $(5.8 \pm 1.0_{\rm stat} \pm 0.2_{\rm syst}) \times 10^{-6}$. This is consistent with theoretical predictions~\cite{Petri:2010ea} and the $(5.8 \pm 0.8) \times 10^{-6}$ PDG value~\cite{PhysRevD.110.030001}. No signal is observed for $\eta\rightarrow e^+ e^-$ , and the upper limit on its  branching fraction at 90\%  CL is set to $2.2\times 10^{-7}$. This represents an improvement over the previous upper limit of $7\times 10^{-7} $ $(CL=90\%)$ from the SND experiment~\cite{SND:2018egu}, providing a stricter experimental constraint on new physics. Recently, theoretical predictions based on a dispersive representation for the $\eta$ transition form factors give $\mathcal{B}(\eta\to\mu^+\mu^-)=(4.54\pm0.04\pm0.02)\times10^{-6}$ and $\mathcal{B}(\eta\to e^+e^-)=(5.37\pm0.04\pm0.02)\times10^{-9}$~\cite{Messerli:2025rnv}. Our measured branching fraction for $\eta\to\mu^+\mu^-$ is in good agreement with this prediction, with the difference being well within $1.2\,\sigma$.

In the future, new experimental facilities, such as the super $\tau$-charm facility (STCF)~\cite{Achasov_2023}, the JLab Eta Factory (JEF)~\cite{Somov:2024jiy}, REDTOP~\cite{Gatto:2019dhj},and the $\eta$ factory of HIAF~\cite{An:2025lws} experiments, will provide larger $\eta$ samples,  enabling improved measurements of $\eta\rightarrow \ell^+ \ell^-$ decays, deeper exploration of the internal structure of the $\eta$ meson, and searches for new physics beyond the SM.


\input{acknowledgement}

\bibliographystyle{apsrev4-2}
\bibliography{main}

\end{document}

%% file: authorlist.tex
M.~Ablikim$^{1}$\BESIIIorcid{0000-0002-3935-619X},
M.~N.~Achasov$^{4,b}$\BESIIIorcid{0000-0002-9400-8622},
P.~Adlarson$^{77}$\BESIIIorcid{0000-0001-6280-3851},
X.~C.~Ai$^{82}$\BESIIIorcid{0000-0003-3856-2415},
R.~Aliberti$^{36}$\BESIIIorcid{0000-0003-3500-4012},
A.~Amoroso$^{76A,76C}$\BESIIIorcid{0000-0002-3095-8610},
Q.~An$^{73,59,\dagger}$,
Y.~Bai$^{58}$\BESIIIorcid{0000-0001-6593-5665},
O.~Bakina$^{37}$\BESIIIorcid{0009-0005-0719-7461},
Y.~Ban$^{47,g}$\BESIIIorcid{0000-0002-1912-0374}
% ...
\\
\vspace{0.2cm}
(BESIII Collaboration)\\
\vspace{0.2cm} {\it
$^{1}$ Institute of High Energy Physics, Beijing 100049, People's Republic of China\\
% ...
$^{4}$ Budker Institute of Nuclear Physics SB RAS (BINP), Novosibirsk 630090, Russia\\
% ...
$^{36}$ Johannes Gutenberg University of Mainz, Johann-Joachim-Becher-Weg 45, D-55099 Mainz, Germany\\
$^{37}$ Joint Institute for Nuclear Research, 141980 Dubna, Moscow region, Russia\\
% ...
$^{47}$ Peking University, Beijing 100871, People's Republic of China\\
% ...
$^{58}$ Southeast University, Nanjing 211100, People's Republic of China\\
$^{59}$ State Key Laboratory of Particle Detection and Electronics, Beijing 100049, Hefei 230026, People's Republic of China\\
% ...
$^{73}$ University of Science and Technology of China, Hefei 230026, People's Republic of China\\
% ...
$^{76}$ University of Turin and INFN, (A)University of Turin, I-10125, Turin, Italy; (B)University of Eastern Piedmont, I-15121, Alessandria, Italy; (C)INFN, I-10125, Turin, Italy\\
$^{77}$ Uppsala University, Box 516, SE-75120 Uppsala, Sweden\\
% ...
$^{82}$ Zhengzhou University, Zhengzhou 450001, People's Republic of China\\

\vspace{0.2cm}
$^{\dagger}$ Deceased\\
% ...
$^{b}$ Also at the Novosibirsk State University, Novosibirsk, 630090, Russia\\
% ...
$^{g}$ Also at State Key Laboratory of Nuclear Physics and Technology, Peking University, Beijing 100871, People's Republic of China\\
% ...

}

%% file: acknowledgement.tex
\section*{Acknowledgements}

The BESIII Collaboration thanks the staff of BEPCII (https://cstr.cn/31109.02.BEPC) and the IHEP computing center for their strong support. This work is supported in part by National Key R\&D Program of China under Contracts Nos. 2025YFA1613900, 2023YFA1606000, 2023YFA1606704; National Natural Science Foundation of China (NSFC) under Contracts Nos. 11635010, 11935015, 11935016, 11935018, 12025502, 12035009, 12035013, 12061131003, 12192260, 12192261, 12192262, 12192263, 12192264, 12192265, 12221005, 12225509, 12235017,12475089 12361141819; the Chinese Academy of Sciences (CAS) Large-Scale Scientific Facility Program; CAS under Contract No. YSBR-101; 100 Talents Program of CAS; The Institute of Nuclear and Particle Physics (INPAC) and Shanghai Key Laboratory for Particle Physics and Cosmology; ERC under Contract No. 758462; German Research Foundation DFG under Contract No. FOR5327; Istituto Nazionale di Fisica Nucleare, Italy; Knut and Alice Wallenberg Foundation under Contracts Nos. 2021.0174, 2021.0299; Ministry of Development of Turkey under Contract No. DPT2006K-120470; National Research Foundation of Korea under Contract No. NRF-2022R1A2C1092335; National Science and Technology fund of Mongolia; Polish National Science Centre under Contract No. 2024/53/B/ST2/00975; STFC (United Kingdom); Swedish Research Council under Contract No. 2019.04595; U. S. Department of Energy under Contract No. DE-FG02-05ER41374.

%% file: main.bib
@article{Young:1967zzb,
    author = "Young, Bing-lin",
    title = "{Rare Decay Modes of the eta Meson}",
    doi = "10.1103/PhysRev.161.1620",
    journal = "Phys. Rev.",
    volume = "161",
    pages = "1620--1630",
    year = "1967"
}

@article{AMETLLER1983301,
title = {The quark triangle: Application to pion and eta decays},
journal = {Nuclear Physics B},
volume = {228},
number = {2},
pages = {301-315},
year = {1983},
issn = {0550-3213},
doi = {https://doi.org/10.1016/0550-3213(83)90326-7},
url = {https://www.sciencedirect.com/science/article/pii/0550321383903267},
author = {Ll. Ametller and L. Bergström and A. Bramon and E. Massó}
}

@article{BABU1982449,
title = {Pseudoscalar electromagnetic form factors: Vector dominance and quantum chromodynamics},
journal = {Phys. Lett. B},
volume = {119},
number = {4},
pages = {449-452},
year = {1982},
issn = {0370-2693},
doi = {https://doi.org/10.1016/0370-2693(82)90710-9},
url = {https://www.sciencedirect.com/science/article/pii/0370269382907109},
author = {K.S. Babu and Ernest Ma}
}

@article{PhysRevD.29.911,
  title = {Soft-momentum quark-model predictions for the pseudoscalar-meson decays into lepton pairs},
  author = {Scadron, M. D. and Visinescu, M.},
  journal = {Phys. Rev. D},
  volume = {29},
  issue = {5},
  pages = {911--915},
  numpages = {0},
  year = {1984},
  month = {Mar},
  publisher = {American Physical Society},
  doi = {10.1103/PhysRevD.29.911},
  url = {https://link.aps.org/doi/10.1103/PhysRevD.29.911}
}

@article{PhysRevD.42.1509,
  title = {$\mathrm{CP}$ violation in $\ensuremath{\eta}$, ${K}_{L}\ensuremath{\rightarrow}\ensuremath{\mu}\overline{\ensuremath{\mu}}$ decays and electric dipole moments of electron and muon},
  author = {Geng, C. Q. and Ng, J. N.},
  journal = {Phys. Rev. D},
  volume = {42},
  issue = {5},
  pages = {1509--1520},
  numpages = {0},
  year = {1990},
  month = {Sep},
  publisher = {American Physical Society},
  doi = {10.1103/PhysRevD.42.1509},
  url = {https://link.aps.org/doi/10.1103/PhysRevD.42.1509}
}

@article{gengLeptonicCPViolation1990,
  title = {Leptonic {{CP}} Violation in Leptoquark Models},
  author = {Geng, C. Q.},
  year = {1990},
  month = jun,
  journal = {Z. Phys. C },
  volume = {48},
  number = {2},
  pages = {279--282},
  issn = {1431-5858},
  doi = {10.1007/BF01554476}
}

@article{Margolis:1992gf,
    author = "Margolis, Bernard and Ng, John and Phipps, Martin and Trottier, Howard D.",
    title = "{Quark model calculation of eta ---\ensuremath{>} lepton+ lepton- to all orders in the bound state relative momentum}",
    reportNumber = "TRI-PP-92-99, MCGILL-92-52",
    doi = "10.1103/PhysRevD.47.1942",
    journal = "Phys. Rev. D",
    volume = "47",
    pages = "1942--1950",
    year = "1993"
}

@article{SAVAGE1992481,
title = {The rare decays π0→e+e−, η→e+e− and η→μ+μ− in chiral perturbation theory},
journal = {Phys. Lett. B},
volume = {291},
number = {4},
pages = {481-483},
year = {1992},
issn = {0370-2693},
doi = {https://doi.org/10.1016/0370-2693(92)91407-Z},
url = {https://www.sciencedirect.com/science/article/pii/037026939291407Z},
author = {Martin J. Savage and Michael Luke and Mark B. Wise}
}

@article{PhysRevLett.80.4633,
  title = {Long-Distance Contributions to the ${\mathit{K}}_{\mathit{L}}\ensuremath{\rightarrow}{\ensuremath{\mu}}^{+}{\ensuremath{\mu}}^{\ensuremath{-}}$ Decay Width},
  author = {G\'omez Dumm, D. and Pich, A.},
  journal = {Phys. Rev. Lett.},
  volume = {80},
  issue = {21},
  pages = {4633--4636},
  numpages = {0},
  year = {1998},
  month = {May},
  publisher = {American Physical Society},
  doi = {10.1103/PhysRevLett.80.4633},
  url = {https://link.aps.org/doi/10.1103/PhysRevLett.80.4633}
}

@article{Ametller:1993we,
    author = "Ametller, L. and Bramon, A. and Masso, E.",
    title = "{The pi0 ---\ensuremath{>} e+ e- and eta ---\ensuremath{>} mu+ mu- decays revisited}",
    reportNumber = "UAB-FT-300-93",
    doi = "10.1103/PhysRevD.48.3388",
    journal = "Phys. Rev. D",
    volume = "48",
    pages = "3388--3391",
    year = "1993"}

@article{Knecht:1999gb,
    author = "Knecht, M. and Peris, S. and Perrottet, M. and de Rafael, E.",
    title = "{Decay of pseudoscalars into lepton pairs and large N(c) QCD}",
    reportNumber = "CPT-99-P-3870, UAB-FT-469",
    doi = "10.1103/PhysRevLett.83.5230",
    journal = "Phys. Rev. Lett.",
    volume = "83",
    pages = "5230--5233",
    year = "1999"
}

@article{Silagadze:2006rt,
    author = "Silagadze, Z. K.",
    title = "{On the two-photon contributions to e+ e- ---\ensuremath{>} eta gamma and e+ e- ---\ensuremath{>} eta-prime gamma}",
    doi = "10.1103/PhysRevD.74.054003",
    journal = "Phys. Rev. D",
    volume = "74",
    pages = "054003",
    year = "2006"
}

@article{Dorokhov:2009xs,
    author = "Dorokhov, A. E. and Ivanov, M. A. and Kovalenko, S. G.",
    title = "{Complete structure dependent analysis of the decay P ---\ensuremath{>} l+ l-}",
    primaryClass = "hep-ph",
    doi = "10.1016/j.physletb.2009.05.033",
    journal = "Phys. Lett. B",
    volume = "677",
    pages = "145--149",
    year = "2009"
}

@mastersthesis{Petri:2010ea,
    author = "Petri, Thimo",
    title = "{Anomalous decays of pseudoscalar mesons}",
    eprint = "1010.2378",
    archivePrefix = "arXiv",
    primaryClass = "nucl-th",
    type = "MS thesis",
    month = "10",
    year = "2010"
}

@article{Masjuan_2016,
   title={η and η ′ decays into lepton pairs},
   volume={2016},
 pages = {108},
ISSN={1029-8479},
   url={http://dx.doi.org/10.1007/JHEP08(2016)108},
   number={8},
   journal={JHEP},
   publisher={Springer Science and Business Media LLC},
   author={Masjuan, Pere and Sanchez-Puertas, Pablo},
   year={2016},
   month=aug }

@article{SND:2018egu,
    author = "Achasov, M. N. and others",
    collaboration = "SND Collaboration",
    title = "{Search for the process $\mathbf{e^+e^- \to\eta}$}",
    primaryClass = "hep-ex",
    doi = "10.1103/PhysRevD.98.052007",
    journal = "Phys. Rev. D",
    volume = "98",
    number = "5",
    pages = "052007",
    year = "2018"
}

@article{PhysRevD.110.030001,
  title = {Review of Particle Physics},
  author = {Navas, S.  and others},
  collaboration = {Particle Data Group Collaboration},
  journal = {Phys. Rev. D},
  volume = {110},
  issue = {3},
  pages = {030001},
  numpages = {5},
  year = {2024},
  month = {Aug},
  publisher = {American Physical Society},
  doi = {10.1103/PhysRevD.110.030001},
  url = {https://link.aps.org/doi/10.1103/PhysRevD.110.030001}
}

@article{PhysRevD.50.92,
  title = {Measurement of the branching ratio for the decay \ensuremath{\eta}\ensuremath{\rightarrow}${\mathrm{\ensuremath{\mu}}}^{+}$${\mathrm{\ensuremath{\mu}}}^{\mathrm{\ensuremath{-}}}$},
  author = {Abegg, R. and others},
  journal = {Phys. Rev. D},
  volume = {50},
  issue = {1},
  pages = {92--103},
  numpages = {0},
  year = {1994},
  month = {Jul},
  publisher = {American Physical Society},
  doi = {10.1103/PhysRevD.50.92},
  url = {https://link.aps.org/doi/10.1103/PhysRevD.50.92}
}

@article{DZHELYADIN1980471,
title = {Study of η→μ+μ− decay},
journal = {Phys. Lett. B},
volume = {97},
number = {3},
pages = {471-472},
year = {1980},
issn = {0370-2693},
doi = {https://doi.org/10.1016/0370-2693(80)90642-5},
url = {https://www.sciencedirect.com/science/article/pii/0370269380906425},
author = {R.I. Dzhelyadin and others}
}

@article{Kang:2023yye,
    author = "Kang, Xiaolin and Ji, Yuyao and Yuan, Xiaoqing and Xiang, Benhou and Zhou, Xiaorong and Peng, Haiping and Huang, Xingtao and Fang, Shuangshi",
    title = "{Novel approach to investigate \ensuremath{\eta} decays via \ensuremath{\eta}'\textrightarrow{}\ensuremath{\pi}\ensuremath{\pi}\ensuremath{\eta}}",
    primaryClass = "hep-ph",
    doi = "10.1103/PhysRevD.108.014038",
    journal = "Phys. Rev. D",
    volume = "108",
    number = "1",
    pages = "014038",
    year = "2023"
}

@article{BESIII:2021cxx,
    author = "Ablikim, M. and others",
    collaboration = "BESIII Collaboration",
    title = "{Number of $J/\psi$ events at BESIII}",
    primaryClass = "hep-ex",
    doi = "10.1088/1674-1137/ac5c2e",
    journal = "Chin. Phys. C",
    volume = "46",
    number = "7",
    pages = "074001",
    year = "2022"
}

@article{ABLIKIM2010345,
title = {Design and construction of the BESIII detector},
journal = {Nucl. Instrum. Meth. A},
volume = {614},
number = {3},
pages = {345-399},
year = {2010},
issn = {0168-9002},
doi = {https://doi.org/10.1016/j.nima.2009.12.050},
url = {https://www.sciencedirect.com/science/article/pii/S0168900209023870},
author = {M. Ablikim  and others},
collaboration = "BESIII  Collaboration"
}

@article{Li:2017jpg,
    author = "Li, Xin and others",
    title = "{Study of MRPC technology for BESIII endcap-TOF upgrade}",
    doi = "10.1007/s41605-017-0014-2",
    journal = "Radiat. Detect. Technol. Meth.",
    volume = "1",
    pages = "13",
    year = "2017"
}

@article{Guo:2017sjt,
    author = "Guo, Ying-Xiao and others",
    title = "{The study of time calibration for upgraded end cap TOF of BESIII}",
    doi = "10.1007/s41605-017-0012-4",
    journal = "Radiat. Detect. Technol. Meth.",
    volume = "1",
    pages = "15",
    year = "2017"
}

@article{CAO2020163053,
title = {Design and construction of the new BESIII endcap Time-of-Flight system with MRPC Technology},
journal = {Nucl. Instrum. Meth. A},
volume = {953},
pages = {163053},
year = {2020},
issn = {0168-9002},
doi = {https://doi.org/10.1016/j.nima.2019.163053},
url = {https://www.sciencedirect.com/science/article/pii/S0168900219314068},
author = {P. Cao and others},
}

@article{AGOSTINELLI2003250,
title = {Geant4—a simulation toolkit},
author = {S. Agostinelli and others},
collaboration = {GEANT4 Collaboration},
journal = {Nucl. Instrum. Meth. A},
volume = {506},
number = {3},
pages = {250-303},
year = {2003},
issn = {0168-9002},
doi = {https://doi.org/10.1016/S0168-9002(03)01368-8},
url = {https://www.sciencedirect.com/science/article/pii/S0168900203013688},

}

@article{JADACH2000260,
title = {The precision Monte Carlo event generator KK for two-fermion final states in e+e− collisions},
journal = {Comput. Phys. Commun.},
volume = {130},
number = {3},
pages = {260-325},
year = {2000},
issn = {0010-4655},
doi = {https://doi.org/10.1016/S0010-4655(00)00048-5},
url = {https://www.sciencedirect.com/science/article/pii/S0010465500000485},
author = {S. Jadach and B.F.L. Ward and Z. Was}
}

@article{PhysRevD.63.113009,
  title = {Coherent exclusive exponentiation for precision Monte Carlo calculations},
  author = {Jadach, S. and Ward, B. F. L. and Was, Z.},
  journal = {Phys. Rev. D},
  volume = {63},
  issue = {11},
  pages = {113009},
  numpages = {65},
  year = {2001},
  month = {May},
  publisher = {American Physical Society},
  doi = {10.1103/PhysRevD.63.113009},
  url = {https://link.aps.org/doi/10.1103/PhysRevD.63.113009}
}

@article{LANGE2001152,
title = {The EvtGen particle decay simulation package},
journal = {Nucl. Instrum. Meth. A},
volume = {462},
number = {1},
pages = {152-155},
year = {2001},
issn = {0168-9002},
doi = {https://doi.org/10.1016/S0168-9002(01)00089-4},
url = {https://www.sciencedirect.com/science/article/pii/S0168900201000894},
author = {David J. Lange},
}

@article{PhysRevD.62.034003,
  title = {Event generator for $J/\ensuremath{\psi}$ and $\ensuremath{\psi}(2S)$ decay},
  author = {Chen, J. C. and Huang, G. S. and Qi, X. R. and Zhang, D. H. and Zhu, Y. S.},
  journal = {Phys. Rev. D},
  volume = {62},
  issue = {3},
  pages = {034003},
  numpages = {8},
  year = {2000},
  month = {Jun},
  publisher = {American Physical Society},
  doi = {10.1103/PhysRevD.62.034003},
  url = {https://link.aps.org/doi/10.1103/PhysRevD.62.034003}
}

@article{BARBERIO1991115,
title = {Photos — a universal Monte Carlo for QED radiative corrections in decays},
journal = {Comput. Phys. Commun.},
volume = {66},
number = {1},
pages = {115-128},
year = {1991},
issn = {0010-4655},
doi = {https://doi.org/10.1016/0010-4655(91)90012-A},
url = {https://www.sciencedirect.com/science/article/pii/001046559190012A},
author = {Elisabetta Barberio and Bob {van Eijk} and Zbigniew Was},
}

@article{Zhang:2012gt,
    author = "Zhang, Zhen-Yu and Qin, Li-Qing and Fang, Shuang-Shi",
    title = "{Event generators for eta/eta-prime rare decays into pi+ pi- l+ l- and e+ e- mu+ mu-}",
    doi = "10.1088/1674-1137/36/10/002",
    journal = "Chin. Phys. C",
    volume = "36",
    pages = "926--931",
    year = "2012"
}

@article{BESIII:2014bgm,
    author = "Ablikim, M. and others",
    collaboration = "BESIII  Collaboration",
    title = "{Observation of $\eta^{\prime}\to\pi^{+}\pi^{-}\pi^{+}\pi^{-}$ and $\eta^{\prime}\to\pi^{+}\pi^{-}\pi^{0}\pi^{0}$}",
    primaryClass = "hep-ex",
    doi = "10.1103/PhysRevLett.112.251801",
    journal = "Phys. Rev. Lett.",
    volume = "112",
    pages = "251801",
    year = "2014"
}

@article{Qin:2017vkw,
    author = "Qin, Nian and Zhang, Zhen-Yu and Fang, Shuang-Shi and Zhou, Xiang and Du, Lin-Lin and Qiao, Hao-Xue",
    title = "{Event generators for $\eta/\eta^{\prime}$ decays at BESIII}",
    primaryClass = "hep-ex",
    doi = "10.1088/1674-1137/42/1/013001",
    journal = "Chin. Phys. C",
    volume = "42",
    number = "1",
    pages = "013001",
    year = "2018"
}

@article{Rolke:2000ij,
    author = "Rolke, Wolfgang A. and Lopez, Angel M.",
    title = "{Confidence intervals and upper bounds for small signals in the presence of background noise}",
    doi = "10.1016/S0168-9002(00)00935-9",
    journal = "Nucl. Instrum. Meth. A",
    volume = "458",
    pages = "745--758",
    year = "2001"
}

@article{liuStudyTrackingEfficiency2025,
  title = {Study of the Tracking Efficiency of Charged Pions at {{BESIII}}},
  author = {Liu, Fang and others},
  year = {2025},
  month = feb,
  journal = "Radiat. Detect. Technol. Meth.",
  volume = "9",
  pages = "390",
url = {https://link.springer.com/article/10.1007/s41605-025-00530-y}
  
}

@article{PhysRevD.87.012002,
  title = {Search for hadronic transition ${\ensuremath{\chi}}_{cJ}\ensuremath{\rightarrow}{\ensuremath{\eta}}_{c}{\ensuremath{\pi}}^{+}{\ensuremath{\pi}}^{\ensuremath{-}}$ and observation of ${\ensuremath{\chi}}_{cJ}\ensuremath{\rightarrow}K\overline{K}\ensuremath{\pi}\ensuremath{\pi}\ensuremath{\pi}$},
 author = "Ablikim, M. and others",
collaboration = "BESIII  Collaboration",
  journal = {Phys. Rev. D},
  volume = {87},
  issue = {1},
  pages = {012002},
  numpages = {13},
  year = {2013},
  month = {Jan},
  publisher = {American Physical Society},
  doi = {10.1103/PhysRevD.87.012002},
  url = {https://link.aps.org/doi/10.1103/PhysRevD.87.012002}
}

@misc{barlow2002systematicerrorsfactsfictions,
      title={Systematic Errors: facts and fictions}, 
      author={Roger Barlow},
      eprint={hep-ex/0207026},
      archivePrefix={arXiv},
      primaryClass={hep-ex},
      url={https://arxiv.org/abs/hep-ex/0207026}, 
}

@article{Achasov_2023,
   title={STCF conceptual design report (Volume 1): Physics &amp; detector},
   volume={19},
   pages = {14701},
   ISSN={2095-0470},
   url={http://dx.doi.org/10.1007/s11467-023-1333-z},
   
   number={1},
   journal={Front. Phys.},
   publisher={China Engineering Science Press Co. Ltd.},
   author={Achasov, M. and others},
   year={2023},
    url={https://journal.hep.com.cn/fop/EN/10.1007/s11467-023-1333-z}, 
   month=nov }

@article{Somov:2024jiy,
    author = "Somov, Alexander",
    title = "{Jlab Eta Factory Experiment in Hall D}",
    doi = "10.22323/1.413.0029",
    journal = "PoS",
    volume = "CD2021",
    pages = "029",
    year = "2024"
}

@misc{Gatto:2019dhj,
    author = "Gatto, Corrado",
    collaboration = "REDTOP Collaboration",
    title = "{The REDTOP experiment}",
    eprint = "1910.08505",
    archivePrefix = "arXiv",
    primaryClass = "physics.ins-det",
    month = "10",
    
}

@misc{An:2025lws,
    author = "An, FengPeng and others",
    title = "{High-Precision Physics Experiments at Huizhou Large-Scale Scientific Facilities}",
    eprint = "2504.21050",
    archivePrefix = "arXiv",
    primaryClass = "hep-ph",
    month = "4",
   
}

@misc{Punzi:2003bu,
    author = "Punzi, Giovanni",
    title = "{Sensitivity of searches for new signals and its optimization}",
    eprint = "physics/0308063",
    archivePrefix = "arXiv",
    journal = "eConf",
    volume = "C030908",
    pages = "MODT002",
    
}

@unpublished{Messerli:2025rnv,
    author = "Messerli, Noah and Hoferichter, Martin and Hoid, Bai-Long and Holz, Simon and Kubis, Bastian",
    title = "{Improved Standard-Model predictions for $\eta^{(\prime)}\to \ell^+ \ell^-$}",
    eprint = "2512.13776",
    archivePrefix = "arXiv",
    primaryClass = "hep-ph",
    month = "12",
    year = "2025"
}
